\documentclass[aps,prl,preprint,groupedaddress]{revtex4}

\usepackage{amssymb}
\begin{document}

\preprint{PRINT-80-0576}

\title{COMPLEX SPINORS AND UNIFIED THEORIES}


\author{Murray Gell-Mann\footnote{Permanent address: 452-48 California Institute of Technology, Pasadena, 
California 91 125 U.S.A. Work supported in part by a grant from the 
Alfred P. Sloan Foundation. }\footnote{Work supported in part by the U.S. Department of Energy under contract No. DEAC-
03-79ER0068}}
\affiliation{CERN\\ 
Geneva, Switzerland }
\author{Pierre Ramond$^\dagger$}
\affiliation{452-48 California Institute of Technology\\ 
Pasadena, California 91125, U.S.A. 
}
\author{Richard Slansky }
\affiliation{Theoretical Division, University of California \\
Los Alamos Scientific Laboratory \\
Los Alamos, NM 87545 U.S.A.
}



{\obeylines
In {\it Supergravity}, 
P. van Nieuwenhuizen \& D.Z . Freedman (eds.) 
{\em North Holland Publishing Company} 1979
}

\maketitle

We were told by Frank Yang in his welcoming speech that supergravity is a phenomenon 
of \underline{theoretical} physics. Why, at this time, is it not more than that? Self-coupled 
extended supergravity, especially for $N = 8$, seems very close to the overall unified theory for which all of us have yearned since the time of Einstein. 
There are no quanta of spin $>2$; there is just one graviton of spin $2$; there are $N$ 
gravitini of spin $3/2$, just right for eating the $N$ Goldstone fermions of spin $1/2$ 
that are needed if $N$-fold supersymmetry is to be violated spontaneously; there are $N(N -1)/ 2$ spin $1$ bosons, perfectly suited to be the gauge bosons for $SO_N$ in the theory with self-coupling. There are $N(N -1) (N-2)/ 6$ spin $1/2$ Majorana particles, 
and with the simplest assignments of charge and colour they include isotopic doublets of quarks and leptons. The theory is highly non-singular in perturbation theory, and the threatened divergence at the level of three loops has not even been demonstrated. The apparently arbitrary cancellation of huge contributions of opposite sign to the cosmological constant (from self-coupling on the one hand and from 
spontaneous violation of supersymmetry on the other) has been phrased in such an elegant way that it may be acceptable. (Of course, if we follow Hawking et al., we may not even need to cancel out the cosmological constant!) 

What is wrong then? Of course the spontaneous violation of $SO_N$ and of supersymmetry 
is not known to happen in the supergravity theory. But what seems much 
worse, the spectrum of elementary particles includes too few spin $1$ and spin $1/2$ objects to agree with the list that we would like to see on the basis of our experience at energies $\lesssim~50$ $GeV$. Of course, looking up at the Planck mass of 
$\sim 2 \times 10^{19}$ $GeV$, we are in a position of greater inferiority than an ant staring up 
at a skyscraper (facing a factor of only $10^6$ or so) and it may not be reasonable to expect that what looks elementary to us should be elementary on a grand scale. 

Nevertheless, we make the comparison and we find that $SO_8\supset  SU_3^C \times U_1\times U_1$ but $SO_8 \not\supset SU_3^c\times  SU_2\times U_1$, so that there is no room for the $X^\pm$ intermediate bosons of 
the charged current weak interaction. Among the spin $1/2$ particles, we have room for at most two flavours of lepton (say $e$ and $\nu_e$) 
and four flavours of quark (say $u$, $d$, $c$ and $s$). Even these numbers may be reduced if we try to locate the Goldstone fermions among the elementary spin $1/2$ particles or if we use the generator of the second $U_1$  in a restrictive way. 

We would then be forced to regard all or most of the known quarks and leptons 
as non-elementary, as well as at least two of the intermediate bosons of the weak 
interaction. The broken Yang-Mills theory of the weak interaction would be only an 
effective gauge theory, not a fundamental one. All this may prove to be the case, 
and we will then have to understand the rather complicated relation existing between 
the elementary particles of the theory and the elementary particles as we perceive them today. 

Various investigators have looked into superconformal supergravity, in which 
one tries to use the full $SU_N$ as a gauge group; such a theory is plagued with particles 
appearing as multiple poles in propagators, involving difficulties with 
negative probabilities or lack of causality. Ignoring these serious difficulties, 
we may ask about the algebraic description of the spin $1/2$ fermions in such a theory.
Apparently they are again connected with third-rank tensor representations, forming 
part of $({\bf N} + {\bf\overline N})_A$ of $SU_N$ (where $A$ means totally antisymmetrized) instead of being 
assigned to $({\bf N})^3_A$ of $SO_N$. 

This tendency to assign the spin $1/2$ fermions to a tensor representation, probably 
a third rank tensor, of $SO_N$ or $SU_N$, exists even in theories having nothing to 
do with supergravity. We may, for example, consider a composite model of quarks and  
leptons, in which they are made up of $N$ kinds of fermionic sub-units. We may think 
of such a scheme in algebraic terms as assigning these sub-units to the representation 
${\bf N}$ of $SO_N$ or $SU_N$ and the quarks and leptons to tensors that are part of $({\bf N})^3$ 
of $SO_N$ or part of $({\bf N }+ {\bf\overline N})^3$ of $SU_N$, provided each known particle is made up of three 
sub-units. (Of course one might use five or a higher odd number and obtain fifth rank tensors and so forth, but three is much simpler.) 

Now what indications come from the attempts to construct a unified Yang-Mills theory? Do they also point to such a third-rank 
tensor for the spin $1/2$ fermions? 

We turn, then, to the program of formulating a broken Yang-Mills theory of 
strong and weak interactions, with an effective energy of unification between $10^{14}$ $GeV$ and the Planck mass. 
This program is only slightly less immodest in conception than the overall unification program of self-coupled supergravity. For the 
sake of expressing all the Yang-Mills coupling constants in terms of a single one, 
a simple group $G$ is employed. (Actually one could use $G \times G, G  \times G \times G$, etc. with 
discrete symmetries connecting the factors, but we shall treat here only the case of a single $G$.) 

The smallest $G$ that has been used is $SU_5$; the known left-handed spin $1/2$ fermions 
are then assigned to three families, each belonging to the reducible representation 
${\bf \bar 5} + {\bf 10}$ of $SU_5$, where ${\bf \bar 5}$ contains $\bar d$, $e^-$, and $\nu_e$ for the lowest family, while ${\bf 10}$ consists 
of $d,~ u,~\bar u$ and $e^+$, The combination ${\bf \bar 5} + {\bf 10}$ is anomaly-free. The violation of 
symmetry takes place in two stages. First the symmetry $SU_5$ is broken down to 
$SU^c_3\times SU_2\times  U_1$ by means of a non-zero vacuum expected value of an operator transforming 
like the adjoint representation ${\bf 24}$, with no direct effect on the fermion 
masses, and then $SU_2\times U_1$ is broken down to $U_1^{e-m}$ by means of operators transforming 
like ${\bf 5} + {\bf \bar 5}$, with perhaps an admixture of ${\bf 45}$ giving the masses of the quarks and 
leptons. The detailed work is done using explicit spinless Higgs boson fields, with 
various constants for mass, for self-coupling and for coupling to fermions, constants 
that must be delicately adjusted to make the masses of the fermions and of the intermediate 
boson for weak interactions tiny with respect to the unification mass. A quantity of roughly similar magnitude, the renormalization-group-invariant mass $\Lambda$ 
of $QCD$, is tiny with respect to the unification mass for a totally different reason, 
namely the smallness of the unified coupling constant, which is $\sim  10^{-2}$ near the 
unification mass, corresponding to the fine structure constant at low energies, and 
is proportional to the reciprocal of the logarithm of $(10^{14}~ GeV)/\Lambda$. Despite some 
successes, which we mention below, the $SU_5$ scheme seems to us a temporary expedient 
rather than a final theory, because of the arbitrariness associated with the Higgs 
bosons and also because the particles and antiparticles among the left-handed 
spin $1/2$ fermions have no relation to each other (i.e., there is no $C$ or $P$ operator for the theory). 

The $SU_5$ scheme has at least two successes: a roughly correct prediction of the 
weak angle $\theta_{\rm w}$  and the prediction that after allowing for renormalization $m_b =m_\tau$,  
which works quite well. The violation of $SU_2\times U_1$  by ${\bf 5} + {\bf \bar 5}$  of $SU_5$ would also give 
$m_s\sim m_\mu$ and $ m_d\sim m_e$   after renormalization. The second of these relations does not 
work but might be subject to large corrections because the quantities are so small; the first might work if the usual estimates of $m_s$ are in error - otherwise some admixture of a ${\bf 45}$ of $SU_5$ has been suggested, along with ${\bf 5} + {\bf \bar 5}$, but affecting mainly 
the two lower families. 

We have studied various complex spinor schemes that reduce to the $SU_5$ system 
after some symmetry violation. work on such schemes has also been done by Georgi 
et al. at Harvard, Susskind and collaborators at Stanford, Wilczek and Zee, G\"ursey 
et al. in the case of $E_6$, and no doubt by many others. Early investigations of 
complex spinor assignments were carried out by Fritzsch and Minkowski. 

First, let us restrict our attention to a single family, say the third one, 
assuming that the $t$ quark exists and that $\nu_{\tau L}$  is nearly massless. We note that  the 
reason that ${\bf \bar 5} + {\bf 10} $ of $SU_5$  is anomaly-free is that the complex spinor representation 
${\bf 16}$ of $SO_{10}$ breaks-up into ${\bf 1} +{\bf  \bar 5} + {\bf 10}$ of $SU_5$, where the singlet can give rise to no 
$SU_5$  anomaly, and all representations of $SO_n$ $( n \ne 6)$ are anomaly-free. 

The 16-dimensional spinor possesses a $C$ symmetry to start with, connecting 
$\tau^-_L \leftrightarrows \tau^+_L $, $b_L\leftrightarrows (\bar b)_L$, etc., and the 16th particle is just the missing $(\bar\nu_\tau)_L$. Symmetry
violations giving fermion masses must correspond to representations contained in the 
symmetrized square of the fermion representation. We note that in $S0_{10}$ we have 
$({\bf 16})^2_s ={\bf 10} + {\bf 126}$  and that with respect to $SU_5$ we have ${\bf 10}\rightarrow {\bf 5}+{\bf \bar 5}$ and ${\bf 126}\rightarrow {\bf 1}+{\bf 45}+{\bf 10}+{\bf \overline{15}}+{\bf \bar 5}+{\bf 50}$.  An operator transforming like the $SU_5$ singlet piece of ${\bf 126}$ 
would break the $SO_{10}$ symmetry down to $SU_5$ and would give a Majorana mass term of 
the form $(\bar\nu_\tau)^2_L+\nu^2_{\tau R}$  to the unobserved neutrino, one that had better be very large 
if the scheme is to work. 

Meanwhile, the ${\bf 10}$ of $SO_{10}$ would give rise to equal Dirac masses for $b$ and $t $
(apart from renormalization) and also to equal Dirac masses for $t$ and $\nu_\tau$. The 
Dirac mass for the neutrino leads directly to a small effective mass $m{\nu_{\tau L}} 
\sim  m_{\rm Dirac}^2/m_{\rm Majorana}$. If $m(\nu_{\tau L}) \approx 1 ~eV$, then neutrinos account for a modest fraction 
of the missing matter in the universe and give a moderate contribution to the gravitational closure of galaxies and clusters of galaxies. 
Putting $m_{\rm Dirac}\approx  m_t \approx  
~ 30~ GeV$ at a guess, the corresponding value of $m_{\rm Majorana}$ would be $\approx 10^{12}~ Gev$. If 
$m_{\rm Majorana}$ is very much smaller than that, the cosmological effects become too large; 
If $ m_{\rm Majorana}$ is much larger, that is harmless, but the cosmological effects become negligible. 

We can examine $SO_{10}$ in a different way by using the decomposition $SO_{10}\supset SO_6\times SO_4$, 
where algebraically $SO_6$ is equivalent to $SU_4$  and $SO_4$  to $SU_2\times   SU_2$. We have, then, 
effectively $SO_{10}\supset SU_2\times SU_2\times SU_4$, where the first $SU_2$ is that of the weak interactions, 
the second one the corresponding $SU_2$ for left-handed antiparticles or right-handed 
particles and $SU_4$  is the generalization of $SU^c_3$ introduced in a different 
connection by Pati and Salam, in which leptons appear as having a fourth colour. 
The representations of $SO_{10}$  then decompose as follows: 
$${\bf 16}\rightarrow ({\bf 2},{\bf 1},{\bf 4})+({\bf 1},{\bf 2},{\bf \bar 4}), \qquad {\bf \overline{16}}\rightarrow ({\bf 2},{\bf 1},{\bf \bar 4})+({\bf 1},{\bf 2},{\bf 4}),\qquad  {\bf 10}\rightarrow ({\bf 2},{\bf 2},{\bf 1})+({\bf 1},{\bf 1},{\bf 6})$$
$${\bf 126}\rightarrow ({\bf 2},{\bf 2},{\bf 15})+({\bf 1},{\bf 1},{\bf 6})+({\bf 3},{\bf 1},{\bf 10})+({\bf 1},{\bf 3},{\bf \overline{10}}).$$
The representations ${\bf 1}$, ${\bf 4}$, ${\bf \bar 4}$, ${\bf 15} $, ${\bf 10}$ and ${\bf \overline{10}}$ of $SU_4$ each contain one colour singlet. 
We see that the Dirac mass term coming from ${\bf 10}$ is just of the form $({\bf 2},{\bf 2},{\bf 1})$, while 
the Majorana mass term for $(\bar\nu_e)_L$ or $\nu_R$ coming from ${\bf 126}$ is a component of $({\bf 1},{\bf 3},{\bf \overline{10}})$. 
We must not use $({\bf 3},{\bf 1},{\bf 10})$, which would introduce an unwanted triplet violation of 
$SU^{\rm weak}_2$ and would give a mass directly to the left-handed neutrino. A possible 
danger is that radiative corrections might give rise to a large or uncontrollable term of that kind anyway, in addition to the term $m^2_{\rm Dirac}/m_{\rm Majorana}$, since the left-handed neutrino Majorana mass is not prohibited by a selection rule. 

Such a selection rule exists in the $SU_5$ scheme, where an ungauged quantity that 
distinguishes ${\bf \bar 5}$ from ${\bf 10}$ and a gauged generator of $SU_5$ are simultaneously violated, 
preserving a linear combination, which is the baryon number minus the lepton number. 
The conservation of this quantity prohibits neutrino mass altogether. Here an 
unwanted massless spinless Goldstone boson is fed to an unwanted massless spin $1$ 
gauge boson to give a massive spin $1$ boson. This trick, which we have studied in 
connection with conserving baryon number (perhaps an obsolete idea now) can be 
applied whenever there is a reducible representation of $G$ for the fermions (or 
even for spinless elementary particles if there are some). 

A further generalization of $SU_5$ for one family might make use of the lowest 
complex spinor representation ${\bf 27}$ of $E_6$, which breaks down to ${\bf 16} + {\bf 10 }+ {\bf 1}$ of $SO_{10}$. 
Here one would have to marry the new $SO_{10}$-singlet neutrino to the unwanted $(\bar\nu)_L$ 
of ${\bf 16}$, allowing them to share a huge Dirac mass, and one would have to do it in such 
 a way as to leave the $\nu_L$  of ${\bf 16}$ with a small mass or none at all. At the same time 
one would have to assign high masses to the members of the ${\bf 10}$ of $SO_{10}$ in the ${\bf 27}$ of 
$E_6$, in order to get them out of the way, leaving just the  fifteen fermions of the $SU_5$ scheme. 

What we have seen from the example of one family is that a complex spinor representation, 
while it involves us in delicate questions of neutrino mass, does permit 
the description of left-handed fermions by a single irreducible representation of 
$G$ and in such a way that the asymmetry between the $SU_2^{\rm weak}$ assignments of particle 
and antiparticle is rather natural, while the whole system possesses an initial 
symmetry $C$ between left-handed particles and left-handed antiparticles, a symmetry 
that interchanges $SU_2^{\rm weak}$  and another $SU_2$. 

It is also clear that in .such a scheme the dimensions of the representations 
that violate the symmetry, for example in the generation of fermion masses, tend to 
be large and that the arbitrary character of the violation scheme employing elementary 
Higgs bosons is strongly emphasized. It seems to us that one must hope for 
a situation in which, somehow, spontaneous symmetry violation is achieved dynamically. 

Although we do not, of course, exclude the existence of some spinless elementary 
fields, provided they are not the arbitrary ones of the elementary Higgs 
boson method, we may look \underline{as an example} at a theory with just gauge bosons and 
elementary spin $1/2$ fields and imagine what hypothetical dynamical spontaneous 
symmetry breaking would be like. 

We would like to point out first that if the (say) left-handed fermion 
representation in such a theory is reducible, then ungauged quantum numbers arise 
that commute with the gauge group. When these are violated spontaneously, that 
necessarily leads to unwanted massless Goldstone bosons unless the trick described 
above is used and global conservation laws result. If all the irreducible representations 
are inequivalent, then such globally conserved quantities are Abelian and 
tolerable, but if there are equivalences among representations, as in the case of 
several families transforming alike, then an ungauged non-Abelian family group 
arises and that would have to be matched with an isomorphic subgroup of $G$ with 
resulting global conservation of a third isomorphic non-Abelian group relating the 
families. That would not agree at all with observation, and we conclude therefore 
that having united each family in an irreducible representation of $SO_{10}$ or $E_6$ we had 
better consider all the fermions as belonging to a single irreducible representation 
of the gauge group $G$.  

This can be done in two different ways. Either we go to a higher-dimensional 
representation of the same group that we used for one family or else we enlarge the 
group and assign the fermions to a relatively low-lying representation of the bigger 
group. In the case of complex spinor representations, we could try, as an example 
of the first approach, the ${\bf 1728}$ of $E_6$, contained in ${\bf 27}\times {\bf 78}$. As examples of the 
second approach, we can take the lowest-dimensional complex spinors of larger groups, 
and the only larger groups possessing such spinors are $SO_{14}$, (lowest dimensional 
spinor ${\bf 64}$), $SO_{18}$ (lowest-dimensional spinor ${\bf 256}$), $SO_{22}$ (lowest-dimensional spinor 
${\bf 1024}$), etc. We have studied both possibilities but we shall describe here the 
case of the lowest spinors of $SO_{4n+2}$.

A great deal of thought has been devoted to the question of what dynamical 
spontaneous symmetry breaking would be like for a theory containing elementary 
fields for gauge bosons and fermions only. Weinberg, Dimopoulos and Susskind, 
and various other theorists have drawn some important conclusions, including the 
following, which we specialize to the case of an irreducible fermion representation. 

Symmetry reduction occurs through ``condensations", that is non-zero vacuum expected 
values of operators that break symmetries. If the symmetry group of the 
kinetic energy is $H$ and if $G_1\subset   G$ and $H_1 \subset H$ are the subgroups left invariant by 
these condensations, then the generators of $G_1$ correspond to exact conservation 
laws and massless gauge bosons, those of $G/G_1$ to massive gauge bosons, those of 
$H_1/G_1$ to modified Goldstone bosons that acquire mass a a result of the gauge coupling 
and those of $(H/G)/(H_1/G_1) $ to approximate conservation laws, broken by the gauge 
coupling . The flavour-non-singlet pseudoscalar mesons would be modified Goldstone 
bosons, and the $PCAC$ condensation $<\bar {q} q>_{vac} \sim \Lambda^3$ presumably occurring in $QCD$ would 
contribute only $\sim ~ e\Lambda$ to the masses of the weak intermediate bosons. If one or more 
additional factors of the exactly conserved strong colour group exist (we prefer to 
call them primed colour, etc.), then these could have higher renormalization-groupvinvariant 
masses $\Lambda'$, etc., and a primed colour group with $\Lambda'\sim  10^3 ~GeV$ could give 
a condensation of fermions possessing primed colour that would account for the weak 
intermediate boson masses. Some of the corresponding pseudoscalar primed mesons 
would serve as effective Higgs bosons to be eaten by these intermediate gauge bosons. 
The mixing between these primed mesons and ordinary pseudoscalar mesons would be 
rather small. There would be no real ultra-violet fermion masses, but only medium-frequency 
or infra-red masses of order $\Lambda$ for quarks, $\Lambda'$ for fermions possessing 
primed colour, and so forth, and then masses obtained by sharing these medium-frequency 
masses through radiative corrections -these last would simulate ultraviolet 
masses up to fairly high energies. 

A great deal of the algebraic behaviour of such symmetry-breaking schemes should 
be simulated by generalized non-linear $\sigma$-models. If those are embedded in linearized 
$\sigma$-models, then one has some connection with the algebraic properties of explicit 
Higgs boson theories. 

An important question is whether the many condensations required for symmetry 
breaking in a unified theory can be explained by the strong long-range interactions 
that appear in the same theory. This is a problem, for example, in connection with 
any condensation leading to Majorana masses for the unwanted neutrinos. 

Now let us return to the notion that $G$ might be $SO_{4n+2}$ with the left-handed 
fermions placed in the $2^{2n}$-dimensional complex spinor representation. Let us consider 
the example of $SO_{18}$, which evidently contains $SO_8 \times SO_{10}$. We can decompose 
the ${\bf 256}$ of $SO_{18}$ as $({\bf 8}_{sp},{\bf \overline{16}})+({\bf 8}_{sp}',{\bf 16})$    of $SO_8\times SO_{10}$, where ${\bf 8}_{sp}$ and ${\bf 8}_{sp}'$ are the two real inequivalent spinors of $SO_8$. We can now write $SO_8\supset   Sp_4\times SU_2$, where the vectorial octet ${\bf 8}_v$ of $SO_8$ can be made to give $({\bf 4},{\bf 2})$ of $Sp_4\times SU_2$ and ${\bf 8}_{sp}'$ of $SO_8$ likewise, 
while ${\bf 8}_{sp}$ of $SO_8$ gives $({\bf 1},{\bf 2}) + ({\bf 2},{\bf 1})$ of $Sp_4\times SU_2$. The ${\bf 256}$ of $SO_{18}$ then 
becomes $({\bf 1} ,{\bf 3},{\bf 16}) + ({\bf 5},{\bf 1},{\bf 16}) + ({\bf 4},{\bf 2},{\bf \overline{16}})$ of $Sp_4\times SU_2 \times SO_{10}$.  If we interpret $Sp_4$ 
as a supplementary factor of the exactly conserved colour group, which becomes 
$SU_3^c\times Sp_4^{c'}$, and $SU_2$ as a gauged family subgroup of $SO_{18}$, we see that the only fundamental 
left-handed fermions without primed colour are three families of 16-dimensional spinors of $SO_{10}$, and we glimpse a possible agreement with experiment. 

We note that $Sp_4^{c'}$, if there were no fermions would have the same renormalization~group behaviour as $SU_3^c$ in lowest order, and we would need a special explanation for its reaching the strong-coupling regime at a much higher mass than $SU_3^c$. The differing 
fermion corrections might make a difference; so might the possibility that as we come down in mass from the unification region $SO_8$ remains undivided over a considerable interval before splitting into $Sp_4\times  SU_2$. 

In the same way, $SO_{14} \supset SO_4\times SO_{10}$ and $SO_4$ is actually algebraically equivalent to 
$Sp_2 \times SU_2$; the ${\bf 64}$ of $SO_{14}$ decomposes into $({\bf 1},{\bf 2},{\bf 16}) + ({\bf 2},{\bf 1},{\bf \overline{16}})$ of $Sp_2\times SU_2\times SO_{10}$ and we would have two families lacking primed colour. Similarly, $SO_{22}\supset   SO_{12}\times SO_{10}$ and $SO_{12}\supset Sp_6\times SU_2$; the ${\bf 1024}$ of $SO_{22}$ decomposes into $({\bf 1},{\bf 4},{\bf 16}) + ({\bf 14},{\bf 2},{\bf 16}) + ({\bf 6},{\bf 3},{\bf \overline{16}}) + ({\bf 14'},{\bf 1},{\bf \overline{16}})$ and we would have four families lacking primed colour. 

As far as representations giving fermion mass are concerned, we have the following situation: 
\begin{eqnarray*}SO_{10}:&&   ({\bf 16})_s^2= ({\bf 10})_A^5({\rm self-dual})+ ({\bf 10})^1={\bf 126}+{\bf 10};\\
SO_{14}:&&   ({\bf 64})_s^2= ({\bf 14})_A^7({\rm self-dual})+ ({\bf 14})^3_A={\bf 1716}+{\bf 364};\\
SO_{18}:&&   ({\bf 256})_s^2= ({\bf 18})_A^9({\rm self-dual})+ ({\bf 18})^1={\bf 24310}+{\bf 8568}+{\bf 18};\\
SO_{22}:&&   ({\bf 1024})_s^2= ({\bf 22})_A^{11}({\rm self-dual})+ ({\bf 22})^7_A={\bf 352,716}+{\bf 170,544}+{\bf 1540};
\end{eqnarray*}
and so forth. It looks in each case as if the Majorana mass term comes from the 
highest-dimensional representation and the Dirac masses of the familiar fermions 
from the next-highest-dimensional one, if such a scheme is to work. The Dirac masses 
then obey an important constraint, which equates a function of the charged lepton 
mass matrix with the same function of the $Q = -1/3$ quark mass matrix. Since in each 
case the matrix is dominated, according to experience, by the highest mass, these 
two masses must be roughly equal, and for three families that explains the relation 
$m_b=m_\tau$ after renormalization. 

The question is, of course, left open as to why the mass matrix for three families is so close to 
$$\begin{pmatrix} {c&0&0\cr 0&0&0\cr 0&0&0}\end{pmatrix}$$
for each kind of particle. With the families described as a triplet of $SU_2$ rather 
than a triplet of $SU_3$, that is rather mysterious, since it corresponds to a miraculous 
compensation of a scalar and a quadrupole term under $SU_2$. Under $SU_3$, of 
course, they would combine to form a ${\bf 6}$ of $SU_3$ and the approximate matrix above 
would correspond to the intervention of the component of ${\bf 6}$ invariant under the 
maximal little group $SU_2$. Unfortunately we are not dealing here with a family $SU_3$. 

The Dirac masses of neutrinos and of $Q = + 2/3$ quarks would obey the same 
relation as that for the charged leptons and $Q = -1/3$ quarks. The Majorana masses of the neutrinos are 
also subject to a constraint if they come from the highest-dimensional representation for the fermion mass. 
Of course, the mass matrix for neutrinos is not easy to detect and at best requires delicate experiments that we shall describe elsewhere. 

In summary, the idea of assigning left-handed spin $1/2$ fermions to a complex 
spinor representation of a gauge group $SO_{4n+2}$ (or conceivably $E_6$) has a number 
of attractive features, although some difficulties as well. As alternatives for 
an irreducible representation, we have, of course, the possibility of a real or 
pseudoreal representation, giving a vector-like theory in which all the known 
fermions must be accompanied by heavy partners that have weak interactions of 
opposite handedness; or a complex representation of a unitary group, which when 
irreducible generally leads to anomalies and thus to divergences, and is also 
rather hard to reconcile with observation. 

If we suppose that the familiar quarks and leptons are really to be assigned 
to a complex spinor representation of a group $SO_{4n+2}$ or $E_6$, can we reconcile that 
idea with the notion that there is some truth in extended supergravity where the 
spin $1/2$ fermions are placed in a third rank antisymmetric tensor representation 
of $SO_N$? 

We have looked, in collaboration with Jon Rosner, for an analogue of supersymmetry that might lead to a theory with assignments like ${\bf 1}$ of 
$E_6$ for $J_z= 2$, ${\bf 27}$ of $E_6$ for $J_z= 3/2$, adjoint ${\bf 78}$ of $E_6$ for $J_z=1$, 
and for $J_z= 1/2$ some representation contained in ${\bf 27}\times {\bf 78}$, like ${\bf 1728}$ of $E_6$. 
We have searched for the same kind of scheme using $SO_{10}$, and we have-even tried non-associative systems in an effort to find something that would work. So far we have had no success. 

It seems likely anyway that if supergravity or some similar future theory is correct, then there must be only an indirect relation between the elementary fields 
of the theory and the particles that appear to us today to be elementary. If the known fermions behave, for a given handedness, like a complex spinor representation 
of $SO_{4n+2}$ or $E_6$, then the relation is not even that of a composite model. All or most of the familiar particles would have to correspond to particle-like solutions of the fundamental equations, with a different algebraic behavior from that of the fundamental fields. 

In this talk we have only sketched the subject of complex spinor representations and related topics. Elsewhere we present a proper account 
of our own work and adequate references to the work of others. 

We have also taken a rather schizophrenic approach, shuttling back and forth between extended supergravity on the one hand and a particular kind of unified 
Yang-Mills theory on the other. The ideas underlying the two approaches have to be compared more carefully. 

\vskip 1cm
\noindent ERRATUM:

\noindent On top of page 8,  the expressions `$H_1/G_1$' and `$(H/G)/(H_1/G_1)$' should be interchanged

\end{document}